# Selective-area van der Waals epitaxy of h-BN/graphene heterostructures via He$^+$ irradiation-induced defect-engineering in 2D substrates

Martin Heilmann[1*], Victor Deinhart[2,3], Abbes Tahraoui[1], Katja Höflich[2,3], and J. Marcelo J. Lopes[1*]

[1] Paul-Drude-Institut für Festkörperelektronik, Leibniz-Institut im Forschungsverbund Berlin e.V., Hausvogteiplatz 5-7, 10117 Berlin, Germany

[2] Ferdinand-Braun Institut, Leibniz-Institut für Höchstfrequenztechnik, Gustav-Kirchhoff-Str. 4, 12489 Berlin, Germany

[3] Helmholtz-Zentrum Berlin für Materialien und Energie GmbH, Hahn-Meitner-Platz 1, 14109 Berlin, Germany

* Corresponding authors e-mail: martinheilmann@gmx.de, lopes@pdi-berlin.de

**The combination of two-dimensional (2D) materials into heterostructures enabled the formation of atomically thin devices with designed properties.[1–3] To achieve a high density, bottom-up integration, the growth of these 2D heterostructures via van der Waals epitaxy (vdWE) is an attractive alternative to the currently mostly employed mechanical transfer, which is still problematic in terms of scaling and reproducibility.[4,5] However, controlling the location of the nuclei formation remains a key challenge in vdWE.[2,6] Here, we use a focused He ion beam for a deterministic placement of defects in graphene substrates, which act as preferential nucleation sites for the growth of insulating, 2D hexagonal boron nitride (h-BN). We demonstrate a mask-free, selective-area vdWE (SAvdWE), where nucleation yield and crystal quality of h-BN is controlled by the**

**ion beam parameter used for the defect formation. Moreover, we show that h-BN grown via SAvdWE has electron tunneling characteristics comparable to those of mechanically transferred layers, thereby lying the foundation for a reliable, high density array fabrication of 2D heterostructures for device integration via defect engineering in 2D substrates.**

Within 2D heterostructures, h-BN plays a central role in two ways: Electronically, its large band-gap enables efficient electronic confinement (as required in ultra-thin layers)[3], while its chemical inertness (i.e. the lack of dangling bonds), together with its insulating nature and high mechanical strength, make it an ideal substrate or encapsulation layer for other 2D materials.[7] Scalable fabrication techniques such as chemical vapor deposition[8,9] or molecular beam epitaxy (MBE)[10,11] were used for the growth of h-BN/graphene heterostructures. The weak out-of-plane interactions between 2D materials leads to preferential nucleation at defects and morphological features in the 2D substrates in vdWE, giving rise to non-uniform growth, and up to now there is no feasible way to control the exact location for the nucleation.[5,11–13] Previous studies on the selective heteroepitaxy on vdW substrates concentrated on the growth of multiple 2D layers on mica or 3D nanostructures on graphene using photo- or electron beam lithography and plasma etching for processing a mask.[14–17] However, these processes have a limited resolution and threaten to contaminate or etch 2D substrates like graphene.

Instead of a mask, we used a focused ion beam (FIB) within a He ion microscope to deliberately create atomic-scale defects in epitaxial graphene (EG) on SiC with perfect position control. The use of EG as a substrate is advantageous for two reasons: (i) the sublimation of Si from SiC leads to a reproducible formation of epitaxially aligned graphene with low intrinsic defect density and (ii) it can be directly synthesized on semi-insulating SiC with no need for a transfer process. According to molecular dynamics simulations, irradiation of 2D materials with ions of noble gases with energies in the keV range results in the formation of defects with control down to single vacancies.[18–20] The irradiation-induced defects (IIDs) in EG exhibit dangling bonds, which act as preferential nucleation sites for atomically thin h-BN grown from elemental B and N sources via plasma-assisted MBE (PAMBE) (Fig. 1a-c). Our mask-free SAvdWE of h-BN is superior to photo- or electron beam lithography, as no residues from the mask threatens to contaminate the surface of the 2D substrate. Ideally, only a single C atom needs to be removed from pristine graphene, as graphene itself serves as a mask through its chemically inert nature.

The processing of EG on semi-insulating SiC substrates results in step edges on the surface with up to 20 nm height (visible in the dark-filed micrograph in Fig. 1d) and several µm wide terraces in between (Supplementary Fig. S1), which are covered with single- (SLG) and bi-layer graphene (BLG). Using a He FIB with an energy of 30 keV the IIDs were written in a pattern, which was aligned to marker structures on the SiC substrate (see method section). The pattern contained 12 areas (Fig. 1d), each consisting of a 19x19 matrix of IIDs in a square grid, written with varying ion numbers per intended defect location (10000, 20000 or 40000 ions), in a distance of either 250 or 500 nm to each other (with two repetitions). Additional cross-shaped markers were written via FIB in-between the different areas and at its corners to facilitate their location during consecutive measurements. Fig. 1e compares the intensities of the D (~1375 cm$^{-1}$) and G (~ 1600 cm$^{-1}$) Raman peaks of graphene in the different areas, with the mean spectra for the denser pattern with varying number of ions used per IID shown in Fig. 1f (the mean spectra of the pattern with 500 nm spacing is shown in Supplementary Fig. S2). For higher ion numbers a small increase of the D/G peak intensity ratio can be observed, indicating a slightly higher defect density.[21] Since the number of IID locations is constant in each pattern, this indicates the formation of more than one defect per intended defect location for higher amounts of He ions. For the G and 2D peak a slight shift can be observed, which we attribute to charge transfer from adsorbates at the IIDs to the graphene after exposure to ambient conditions.[22] Mapping the D peak intensity, we can distinguish the structured area of 19x19 defects in a dense pattern (with 40000 ions per IID) from the surrounding pristine EG (inset of Fig. 1f).

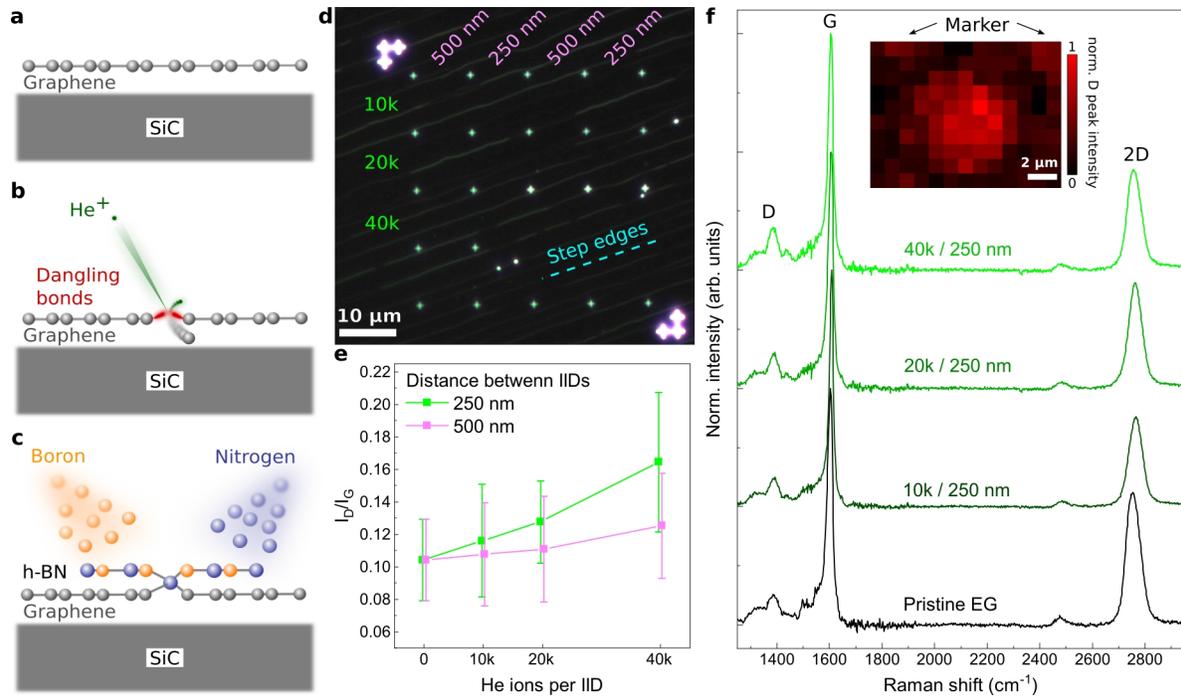

**Fig. 1 | Defect-engineering in EG on SiC via He FIB**. **a-c** Schematic representation of the SAvdWE process. **a**, Pristine EG fabricated via Si sublimation from SiC substrates. **b**, Deterministic placement of IIDs in EG via He FIB. **c**, Growth of h-BN via PAMBE, where dangling bonds at the defects act as preferential nucleation sites. For simplicity the C-rich interface layer to the SiC was neglected in these images. **d**, Dark-field micrograph from structured area showing the 4x3 matrix of the FIB pattern, each containing 19x19 IIDs, written in a square grid with varying distance to each other and different number of He ions as indicated by magenta and green annotations, respectively. SiC step edges run parallel to the cyan dotted line. Additional cross-shaped markers were written between the different areas via He FIB. **e**, **f**, Raman characterization of IID pattern. **e**, Average ratio between D and G peak intensity for 36 spectra recorded in each area in the FIB pattern (**d**). The two different data sets are laterally offset for clarity. The error bars represent one standard deviation. **f**, Comparison of the mean spectra of patterns with higher IID density written with varying number of He ions per defect. The spectra were normalized with respect to the G peak and vertically shifted for clarity. *Inset:* Raman map of the D peak intensity marking the structured area containing 19x19 IIDs, each written with 40000 He ions in a distance of 250 nm to each other, in between markers (indicated by black arrows).

Following the IID preparation, the substrates were introduced to the MBE system for the SAvdWE of h-BN islands (see Method section). Fig. 2a presents an atomic force microscopy (AFM) image of an area of 19x19 IIDs, each written with 20000 ions, in a distance of 250 nm (areas with 500 nm distance can be seen in Supplementary Fig. S2). Areas with a rougher surface can be distinguished from smoother areas close to the step edges, which we attribute to overgrown SLG and BLG, respectively, as shown in previous publications.[11,12] In EG, SLG is known to have a different chemical reactivity then BLG, due to an atomic corrugation of the C-rich and partially sp$^3$-bonded interface layer to SiC, which is screened by an additional graphene layer in BLG, leading to the different growth behavior.[23,24] As the focus of the current study lies on the SAvdWE of h-BN islands at IIDs, we will concentrate in the discussion on the growth of h-BN at IIDs on BLG, where the influence of the interface layer, which is specific for EG on SiC, is negligible. The corresponding conductive AFM (cAFM) image of the same area (Fig. 2b) shows a clear contrast between graphene and h-BN islands, which appear as insulating areas around the intended locations for the IIDs within the markers. In the detailed AFM image shown in Fig. 2c, a mean diameter of 150 nm of the h-BN islands can be observed on the BLG areas. Their average height of 0.34 nm (see also profile along blue dotted line in Fig. 2e) corresponds well to previously reported values for h-BN islands grown on graphene and graphite.[10,12] The distance of ~ 250 nm between the islands and their square arrangement precisely follow the FIB pattern. For single crystalline h-BN islands either a triangular or hexagonal shape would have been expected[25], but the islands here show a non-regular shape. Additional h-BN layers can be observed close to the center of the h-BN islands, together with ~ 2 nm high nanoparticles (NPs), similar to the formation reported from naturally occurring defects in EG.[12] Fig. 2d shows the corresponding cAFM image of the same area, where a clear contrast is observed not only between graphene and h-BN islands, but also within the islands, where the central area with the additional layers and NPs shows no tunneling current at a bias of 100 mV.

The Raman spectra of these samples shown in Fig. 2f reveal a substantially increased intensity at ~ 1360 cm$^{-1}$, where the $E_{2g}$ peak of h-BN is expected.[26] Similar to previous studies a shift of the G and 2D peak can also be observed, which can be attributed to a change in strain and doping of graphene.[11]

In aligned h-BN/graphene heterostructures fabricated via mechanical exfoliation a strained-induced broadening of the 2D peak (as we observe it here) was also reported.[27] Comparing the mean spectra of the patterned areas with non-patterned areas, a slightly higher intensity of the h-BN related Raman peak can be observed, which we attribute to the denser nucleation of h-BN.

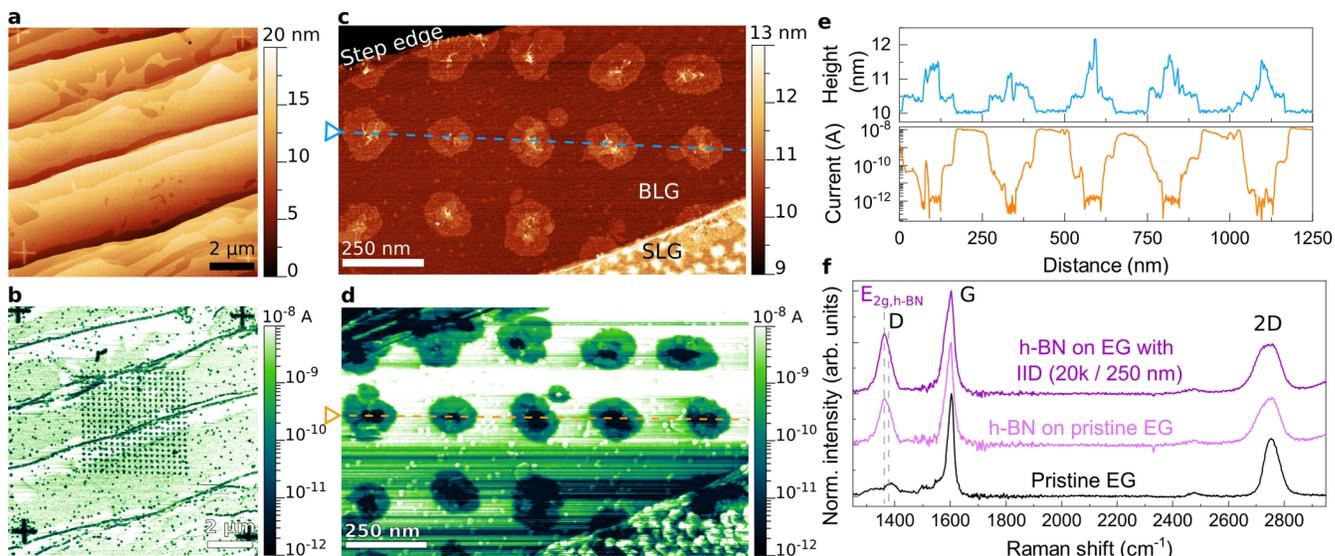

**Fig. 2 | h-BN islands grown via SAvdWE for 300 min**. **a**, **b** AFM image of an area with 19x19 defects written with 20 000 ions per IID in a distance of 250 nm to each other between the cross-shaped markers, with the corresponding cAFM image with a bias of 100 mV applied to graphene showing the position of the insulating h-BN islands in (**b**). **c**, Detailed AFM image of h-BN islands forming close to a step edge on BLG with the corresponding cAFM image shown in (**d**) with a sample bias of 200 mV. **e**, height profile (upper graph) and corresponding current (lower graph) from the blue and orange dotted lines in (**c**) and (**d**), respectively. **f**, Comparison of Raman spectra from pristine graphene with spectra from h-BN islands grown on non-structured areas as well as h-BN islands grown via SAvdWE. The spectra were normalized with regard to the G peak and are vertical shifted for clarity.

In Fig. 3a and 3b we compare the cAFM images of the areas structured with 40000 and 10000 ions per IID (corresponding AFM images are shown in Supplementary Fig. S3). For the higher amount of He ions in every location of an IID on the terraces a h-BN island can be found, while for fewer ions only in 86 % of the IIDs a h-BN island nucleated (with most islands missing on BLG, see also Supplementary Fig. S3). The h-BN islands nucleating at IIDs written with 40000 ions per IID have a diameter of ~ 165 nm and a non-hexagonal shape (see Fig. 3c). Furthermore, the islands contain additional layers and a cluster of NPs in the center, indicating a similar formation as in the area studied before (see Fig. 2).

According to molecular dynamics simulations the sputtering rate of C atoms from graphene upon He ion irradiation with an energy of 30 keV is expected to be very low.[20] However, the defect formation in 2D materials is not only determined by the ion mass and its energy, but is also affected by the interactions with the substrate, e.g. by back-scattered ions or atoms sputtered from the substrate.[20] Hence, despite the low sputtering rate, by using enough He ions the FIB could remove more than only one C atom per intended defect location (as indicated by the Raman spectra shown in Fig. 1). At such laterally extended defects we assume multiple nucleation points for h-BN, potentially with different in-plane rotation, forming polycrystalline islands with NPs in their center. Based on this assumption, fewer ions with the same ion energies should lead to fewer C atoms being knocked out, resulting at best in a single vacancy, which in turn may allow one h-BN domain only to nucleate as a single crystal.

The morphology of the h-BN forming at IIDs written with 10000 ions is shown in Fig. 3d. Several mono-layer thick h-BN islands on BLG present a hexagonal shape, with well defined edges and without wrinkles. Their mean diameter amounts to ~ 135 nm and they have a common alignment among each other (indicated by green dashed arrows). Furthermore, considering the well-known epitaxial relationship between EG and SiC (EG[10$\bar{1}$0] // SiC[11$\bar{2}$0]) and the orientation of the surface step edges (see Fig. 1d), the results suggest that the edges of the h-BN islands parallel to the green arrows follow the [11$\bar{2}$0] crystalline direction in EG. Unlike the non-hexagonal islands (see cAFM image in Fig. 3e), these oriented islands exhibit a Moiré pattern in their conductivity (Fig. 3f). The detailed cAFM image of an exemplary island (Fig. 3g) and the profile along the orange dotted line (Fig. 3h) reveal a periodicity of the Moiré pattern of ~ 14 nm, which is expected for epitaxially aligned h-BN on graphene, with EG[10$\bar{1}$0] // h-BN[10$\bar{1}$0].[28] A similar periodicity can be observed in all hexagonal-shaped islands. Hence, single-crystalline and epitaxially aligned h-BN islands can be grown via SAvdWE by defect engineering in the EG substrate using a He FIB.

Furthermore, unlike in the non-hexagonal shaped islands, the centers of the epitaxially aligned islands do not always coincide with the location of the IIDs. This could indicate that smaller defects (single vacancies), written with 10000 He ions per IID, may heal out or move during heating of the substrate to the growth temperature.[29] In contrast, we expect that using 20000 or 40000 He ions per defect location resulted in the formation of laterally extended defects, which are more stable upon thermal treatment. Further studies will elucidate the nature of the IIDs and develop protocols for their stabilization.

The NPs on the well aligned islands shown in Fig. 3d also got partially removed during the cAFM measurements (see Supplementary Fig. S4), indicating their relatively weak bonding to the h-BN as compared to the clustered NPs observed in the islands nucleating at IIDs written with 20000 and 40000 He ions. This demonstrates that the deterministic placement of defects in the 2D substrate does not only affect the growth of the first 2D layer, but also the formation of additional layers.

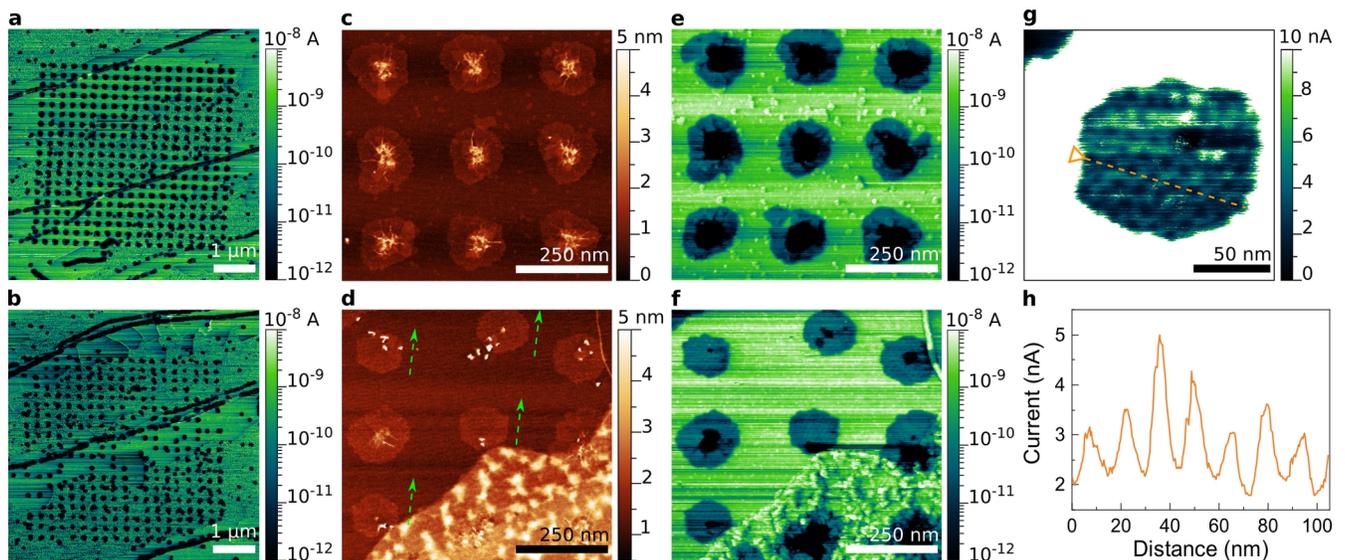

**Fig. 3 | SAvdWE of h-BN at IIDs written with varying number of He ions**. **a,b**, cAFM images of areas with 19x19 overgrown IIDs in a distance of 250 nm to each other, written with (**a**) 40000 and (**b**) 10000 He ions per IID. A bias of 100 mV was applied to graphene during the measurements. **c-f**, Detailed AFM images of h-BN islands forming close to a step edge on BLG patterned with (**c**) 40000 and (**d**) 10000 He ion per IID with corresponding cAFM images at a bias of 200 mV in (**e**) and (**f**), respectively. The green dashed arrows in (**d**) act as a guide to they eye depicting the common orientation among the hexagonal shaped h-BN islands. **g**, cAFM image of a single h-BN island with linear scale. **h**, Profile along orange dotted line in (**g**) demonstrates a Moiré pattern of ~ 14 nm in the conductivity, evidencing an epitaxial alignment between the island and the EG.

The vertical and lateral formation of the h-BN islands from the IIDs via SAvdWE was studied further by using an increased deposition rate of B and longer growth time to form a coalesced h-BN layer between the IIDs. The AFM image in Fig. 4a depicts an overgrown surface area in BLG, which was pat-terned with 20000 ions per defect location, in a distance of 250 nm to each other. Inside the patterned area a completely coalesced h-BN mono-layer formed due to the higher nucleation yield at the IIDs, while outside of the pattern BLG remains still largely uncovered (see Supplementary Fig. S5).

Apart from ~ 8 nm high NPs some additional h-BN layers with diameters of up to 100 nm formed close to some IIDs, which have a relatively lower tunneling current, as shown in the cAFM image in Fig. 4b (some NPs got removed during the cAFM scan). Furthermore, between some differently oriented domains we observe lower tunneling currents at the grain boundaries, where the morphology shows a network of wrinkles. This suggests the formation of overlapping h-BN layers upon the coalescence of the different domains in these areas, similar to observations in h-BN grown via catalytic chemical vapor deposition on metal foils.[30] In other areas we also observe the formation of atomically smooth grain boundaries (see Fig. 4c), where differently oriented grains merged seamlessly. No increased tunneling current could be detected from either of these grain boundaries, which is a key aspect if such h-BN layers are to be employed as tunneling barriers.

In comparison to standard growth, another advantage of SAvdWE is the potential of controlling grain boundary location by IID pattern design. Furthermore, neither the overlapping nor the seamless grain boundaries induced preferential formation of additional h-BN layers. Instead, additional layers can be found in the vicinity of the IIDs. This demonstrates that the artificially created defects can act as preferential nucleation points for the formation of additional h-BN layers on top of the first coalesced one. Thereby it appears plausible, that also other 2D materials could be connected to the IIDs upon completion of the first 2D layer, allowing more complex 2D heterostructures to form via SAvdWE. For the future device integration of such heterostructures, the active region could then be designed such, that it only incorporates the area in between the IIDs, where the heterostructures grow smoothly.

Fig. 4d shows representative I-V-curves from different locations on the samples with varying numbers of h-BN layers (the measurement configuration is depicted in the inset). The tunnel conductance shown in Fig. 4e was extracted from the linear regime of these I-V curves and the observed exponential decay with increasing layer number is expected for direct tunneling of electrons. A similar tunneling behavior was observed for h-BN flakes, exfoliated from high-quality h-BN bulk crystals, showing that the electronic quality of the h-BN grown via SAvdWE can compete with heterostructures fabricated via mechanical transfer.[31,32]

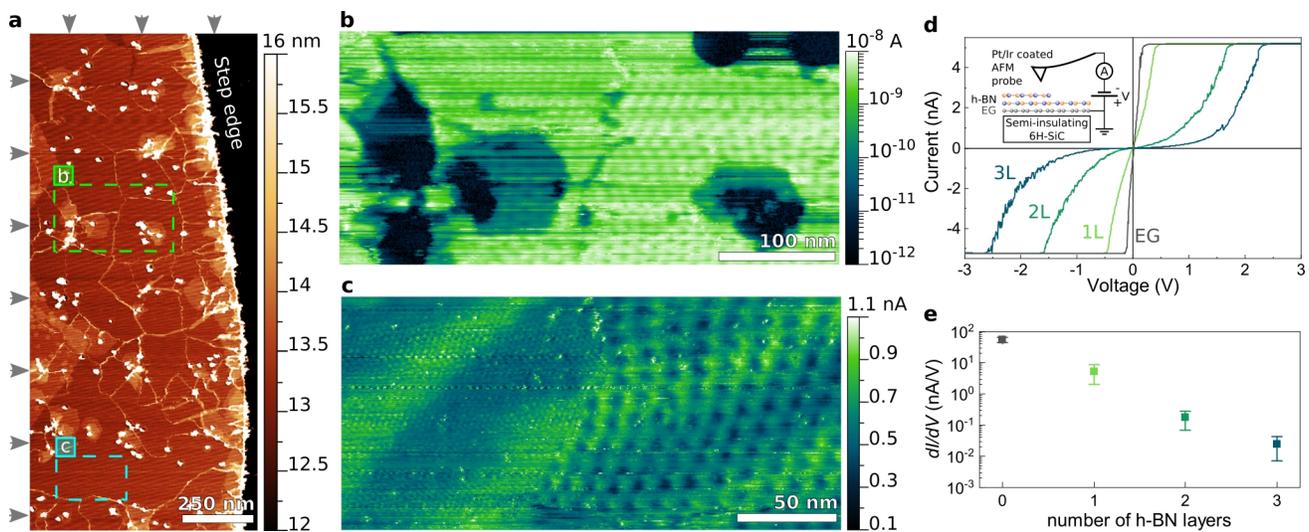

**Fig. 4 | Coalesced h-BN layers between IIDs in EG**. **a**, AFM images of an area close to a step edge with IIDs in a distance of 250 nm written with 20000 He ions per IID, overgrown with h-BN for 480 min. Rows and columns of the IIDs with 250 nm spacing are marked by gray arrows. **b**, cAFM image of the green framed region in (**a**), where multiple h-BN layers can be observed close to IIDs. **c**, cAFM image of an atomically smooth area in between 4 IIDs as marked by the cyan colored frame in (**a**), demonstrating a grain boundary between two differently oriented domains, as evidenced by different periods of the Moiré patterns (3 and 14 nm). A bias of 100 mV was applied to graphene in (**b**) and (**c**). **d**, **e**, Vertical transport measurements of h-BN tunnel barriers, with (**d**) representative I-V curves of mono- (1L), bi-(2L) and tri-layer h-BN (3L) (*inset*: measurement configuration) and (**e**) the log. scale tunnel conductance extracted from the linear regime of the I-V curves. The error bars represent one standard deviation.

In conclusion, we demonstrated a mask-free selective area epitaxy of two-dimensional h-BN on graphene using a deterministic placement of defects via He FIB. The SAvdWE allowed an unprecedented control over the formation of epitaxially aligned and single-crystalline 2D heterostructures via a bottom-up approach. Additionally to the nucleation location, the lateral formation of h-BN on EG was controlled via the growth parameter during the growth process, resulting in a 2D h-BN layer with tunneling characteristics identical to exfoliated material. This approach is not only limited to h-BN on EG, but should be applicable to other 2D materials, which have been shown to preferentially nucleate at defects in vdW substrates.[13,33] Therefore, the SAvdWE contributes to precise and scalable fabrication of 2D heterostructures by establishing a reliable processing route for the site-selective, epitaxial bottom-up manufacturing with the potential for a reproducible fabrication of nano-devices based on 2D materials.

**Acknowledgments**

The authors would like to acknowledge fruitful discussions with Dr. Arkady Krasheninnikov on the irradiation induced defect formation in 2D materials. We would also like to thank Werner Seidel for his help in the preparation of the patterned SiC substrates. Furthermore, we are grateful to Carsten Stemmler, Hans-Peter Schönherr and Claudia Herrmann, for their dedicated maintenance of the PAMBE system and appreciate the critical reading of the manuscript by Dr. Alberto Hernández-Mínguez. The He ion beam patterning was performed in the Corelab Correlative Microscopy and Spectroscopy at Helmholtz-Zentrum Berlin.

**Methods**

*Substrate preparation.* The markers for the alignment of the FIB pattern were prepared on the 1x1 cm² large 6H-SiC substrates with a miscut of ~ 0.2° using photolithography, followed by reactive ion etching using $SF_6$. After chemically cleaning, the SiC(0001) was etched for 15 min at 1450 °C in forming gas (5 at % $H_2$ in Ar) using a flux of 500 sccm at a pressure of 900 mbar. In a second step EG formed on the etched SiC via sublimation of Si for 30 min at 1650 °C in Ar at a pressure of 900 mbar. Further information on the etching and sublimation process can be found elsewhere.[34,35] For the thermal cou-pling during the growth of h-BN via MBE the backside of the SiC substrate was coated with 1 µm of Ti via electron beam evaporation.

*He FIB patterning.* The samples were loaded to the FIB system one day in advance to the actual patterning to allow for desorption of adhering species during pumping over night. The background pressure during patterning was around $1 \times 10^{-7}$ mbar. After adjusting column and beam all stage moves and patterning tasks were carried out without imaging. Hence, unintended ion beam irradiation of the patterned regions could be avoided. The pattern was designed as described in the discussion of Fig. 1, with varying dwell times and distance between IIDs. The IIDs were prepared using a 30 keV He ion beam at an ion current of 1.8 pA. With the settings for the FIB used in this study, the dwell times were varied between of 4, 2 and 1 ms correspond to approximately 40000, 20000 and 10000 ions per intended defect location, respectively. The appropriate ion number for creating defects which initiate nucleation was determined on test samples, were the ion dose was varied over several orders of magnitude covering the range from one ion per IID up to 130 000 ions. A typical test pattern for this purpose contained 6 individual IIDs as well as IIDs written along 6 lines, each 1 µm long, to ensure sufficient statistics (see supplementary Figure S6).

*PAMBE of h-BN.* After loading the patterned EG samples into the MBE system, they were thermally cleaned in a preparation chamber for 30 min in UHV ($1 \times 10^{-9}$ mbar) at 400 °C. Consecutively, the h-BN islands (coalesced h-BN layers) were grown for 300 min (480 min) at a substrate temperature of 850 °C using a high temperature effusion cell operated at ~ 1850 °C (1900 °C) as a B source, whereas N was supplied using an Addon RFX450 RF plasma source operated at 350 W and a $N_2$ flux of 0.2 sccm. The temperature was monitored using an optical pyrometer, and the heating and cooling rates were ~ 20 °C/min. Further information on the growth behavior of h-BN on EG at similar conditions can be found elsewhere.[11,12]

*Micro-Raman spectroscopy.* Raman measurements were conducted in back-scattering configuration using an excitation wavelength of 473 nm, with the laser focused on the sample using a 100x objective with a NA of 0.9, resulting in a spot size of ~ 1 µm. The reflected light was analyzed using a grating with 600 gr. mm$^{-1}$ and a liquid $N_2$ cooled Si CCD camera. Raman spectra were recorded for each pattern in a 5x5 µm$^2$ map containing 36 measurement spots (each with 2x60 sec integration time) on the patterned sample with and without h-BN.

*AFM and cAFM.* The morphology of the samples was monitored using AFM in intermittent contact mode (tip diameter, 2 nm; spring constant, 0.4 N/m) with the cantilever deflection used as feedback.[36] AFM scans were aligned with regard to the marker in the FIB pattern (e.g., see Fig. 1e and Supplementary Fig. S1). The cAFM measurements were conducted in contact mode at ambient conditions using Pt/Ir coated tips (tip diameter, 25 nm; spring constant, 0.4 N/m). A bias between 100 and 200 mV was applied to the EG on the semi-insulating SiC as described in the text, and the tunneling current between graphene and AFM tip was measured at a constant force of ~ 10 nN. I-V curves were recorded in $N_2$ atmosphere at 144 locations in each of the areas shown in Fig. 4a and 4b with 5 repetitions per location.

# Supplementary information:

# Selective-area van der Waals epitaxy of h-BN/graphene heterostructures via He⁺ irradiation-induced defect-engineering in 2D substrates

Martin Heilmann[1], Victor Deinhart[2,3], Abbes Tahraoui[1], Katja Höflich[2,3], and J. Marcelo J. Lopes[1]

[1] Paul-Drude-Institut für Festkörperelektronik, Leibniz-Institut im Forschungsverbund Berlin e.V., Hausvogteiplatz 5-7, 10117 Berlin, Germany

[2] Ferdinand-Braun Institut, Leibniz-Institut für Höchstfrequenztechnik, Gustav-Kirchhoff-Str. 4, 12489 Berlin, Germany

[3] Helmholtz-Zentrum Berlin für Materialien und Energie GmbH, Hahn-Meitner-Platz 1, 14109 Berlin, Germany

For locating the areas structured with a He focused ion beam (FIB) the 10x10 mm² SiC samples were first structured with markers via photolithography and reactive ion etching using $SF_6$. After the processing of epitaxial graphene (EG) the markers were still visible (see dark field micrograph in Fig. S1a) and the FIB pattern (marked by the magenta framed square, and as shown in Fig. 1d in the main text) was aligned to them. Fig. S1b shows the morphology of one structured area after the growth of h-BN for 300 min, as measured by atomic force microscopy (AFM). On the terraces rougher areas can be distinguished from smoother areas close to the step edges, which we attribute to overgrown single- (SLG) and bi-layer graphene (BLG). Previous studies already demonstrated faster growth rates of disordered h-BN on SLG due to an atomic corrugation of the SLG, induced by the closer proximity to the interface layer to SiC, which is screened by an additional graphene layer in BLG.[1]

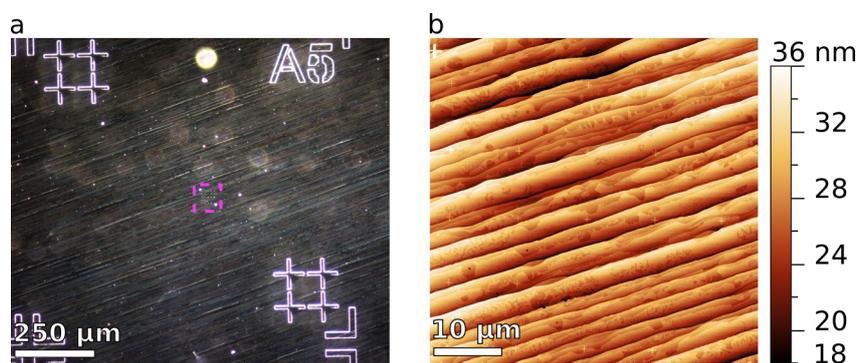

**Fig. 1a**, Dark-field micrograph of a marker on the substrate with the FIB pattern marked by the magenta dotted square. **b**, AFM image of the patterned area.

Fig. S2 compares the mean Raman spectra for the FIB pattern with 500 nm distance between the irradiation induced defects (IIDs) written with varying numbers of He ions. Unlike in pattern with a distance of 250 nm between the IIDs shown in Fig. 1 of the main text, only in the case of 40000 ions per IID a slight increase in the D peak intensity was observed, due to the lower density of defects in this pattern. The AFM image in Fig. S2b shows a sample area with IIDs written with 20000 ions per defect after the growth for 300 minutes at 850 °C. The corresponding conductive AFM (cAFM) image shown in Fig. S2c shows the position of the h-BN islands within the marked area.

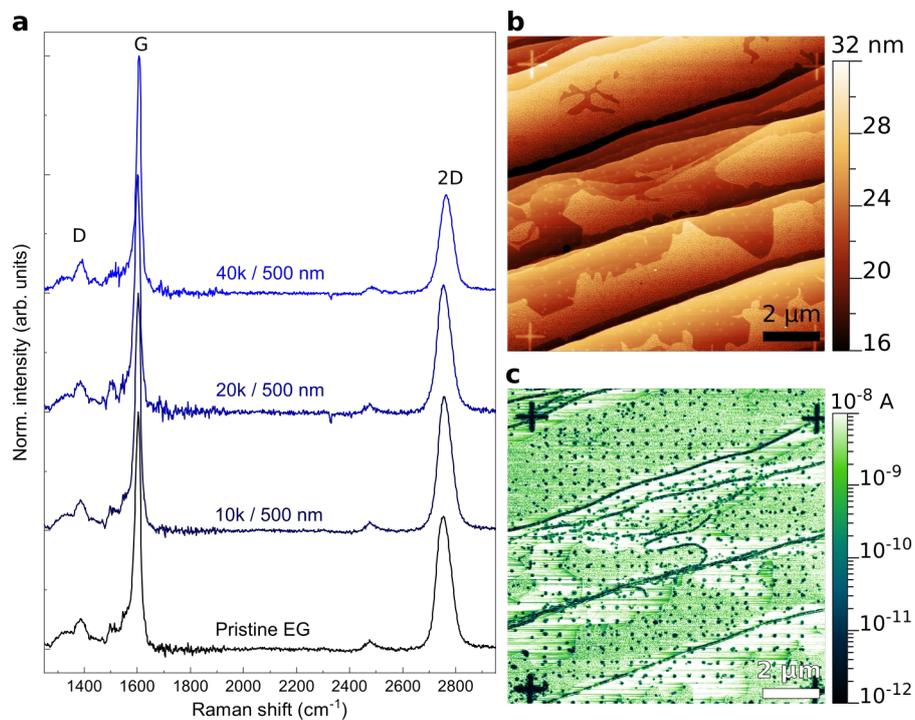

**Fig. S2a**, Comparison of the mean Raman spectra of patterns with 500 nm distance between the IIDs density written with varying number of He ions per defect. The spectra were normalized with respect to the G peak and vertically shifted for clarity. **b**, AFM images of an area with 19x19 defects written with 20 000 ions per IID in a distance of 500 nm to each other between the cross-shaped markers, with the corresponding cAFM image at a bias of 100 mV applied to graphene showing the position of the insulating h-BN islands in (**c**).

Fig. S3 shows the morphology and cAFM images of areas structured with different amounts of He ions per defect location after the growth of h-BN for 300 min. A difference in nucleation yield is apparent, where for fewer He ions only 86 % of the IIDs on the terraces resulted in a nucleation of h-BN.

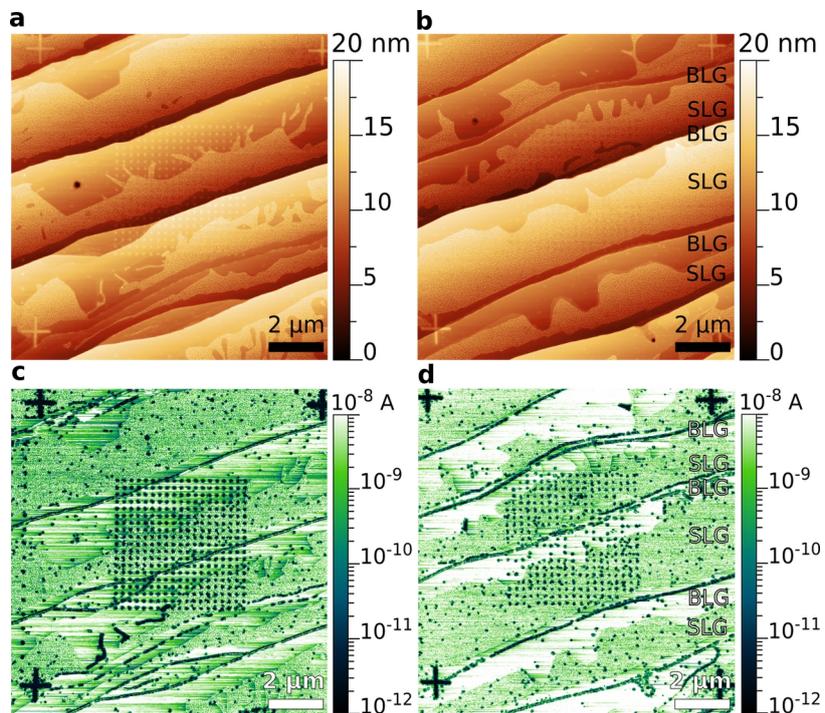

**Fig. S3** AFM images of patterned areas with (**a**) 40000 and (**b**) 10000 ions per IID after the growth of h-BN for 300 min, with the cAFM images in (**c**) and (**d**), respectively.

Due to the imaging in contact mode during cAFM measurements some nanoparticles (NPs) were removed from the epitaxially aligned h-BN islands nucleating at IIDs written with 10000 He ions each, as shown in Fig S4.

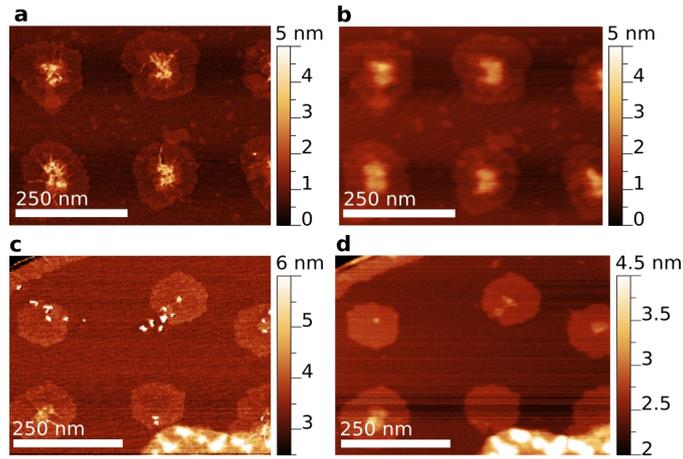

**Fig. S4** AFM images of h-BN islands in patterned areas with 40000 ions per IID (**a**) before and (**b**) after cAFM measurements, show no change in morphology. AFM image of h-BN islands nucleating at IIDs written with 10000 He ions (**c**) before and (**d**) after cAFM measurements show the removal of some NPs.

Fig. S5a shows an AFM image of the surface of a patterned area with 20000 He ions per IID (with a distance of 250 nm to each other) after the overgrowth of h-BN for 480 min with increased B deposition rate, with the corresponding cAFM image shown in Fig. S5b. Due to their atomic corrugation the single-layer graphene (SLG) areas are completely overgrown with multiple, disordered h-BN layers[1] and appear mostly insulating. On bi-layer graphene (BLG) only in the patterned area one completely coalesced h-BN layer formed, while in the non-patterned BLG areas h-BN is mostly growing at step edges and wrinkles, leaving large parts of graphene uncovered.

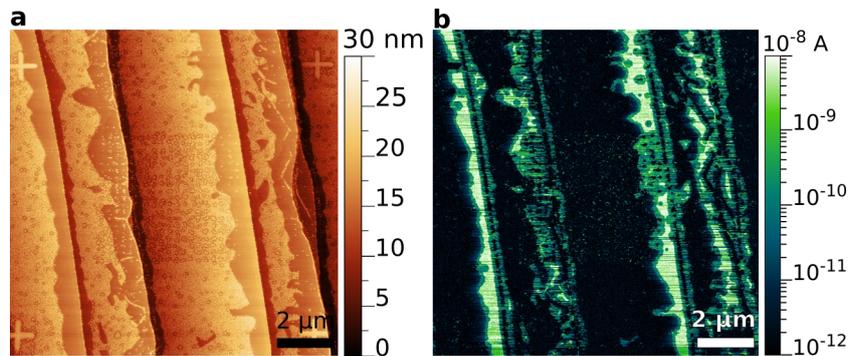

**Fig. S5a**, AFM image of patterned areas with 20000 ions per IID after the growth of h-BN for 480 min at increased B deposition rate, with the corresponding cAFM image in (**b**).

The appropriate ion number for creating defects which initiate nucleation was determined on test samples, were the number of ions was varied over several orders of magnitude covering the range from one ion per IID up to 130000 ions. A typical test pattern for this purpose contained 6 individual IIDs (in a distance of 500 nm) as well as overlapping IIDs written along 6 lines, each 500 nm long. The SEM image show in Fig. S6a shows a patterned area, where the highest numbers of ions were used. The dark and bright contrast in the sample are attributed to SLG and BLG. The individual IIDs are visible with a relatively higher contrast, while the IIDs written along lines appear darker as the background. To study the nucleation yield, the test structures were loaded into the MBE and h-BN was grown for 300 minutes at 850 °C with the same sources and settings as shown in the main text. Two repetitions of these structures were then analyzed using AFM and cAFM. Fig. S6b summarizes the results for the individual IIDs, where below ~ 1000 ions per IID no preferential nucleation could be observed at the defects, with only in few cases where an island probably grew at a location of an IID by chance (see AFM image in Fig. S6c with corresponding cAFM image in Fig. S6d). Above 10000 He ions in each location of an IID an insulating h-BN island could be found leading to the nucleation yield of 100 % (see AFM image in Fig. S6e and S6g with corresponding cAFM image in Fig. S6f and S6h, respectively). The difference to the nucleation yield of 86 % at 10000 He ions per IID observed in the structures shown in the main text is attributed to the lower amount of IIDs written in the test structures.

The intention behind the overlapping IIDs written along lines was to use them as markers to locate the patterned areas after the growth. Here we decided to use 70000 ions per IID, where after the growth lines with a heights above 50 nm could be observed (see Fig. S6g). In between the patterned areas additional cross-shaped markers were written (e.g. as shown in Fig. S3) using 2000 ions per overlapping IID. After the growth the height of these cross-shaped markers amounted to ~ 2 nm.

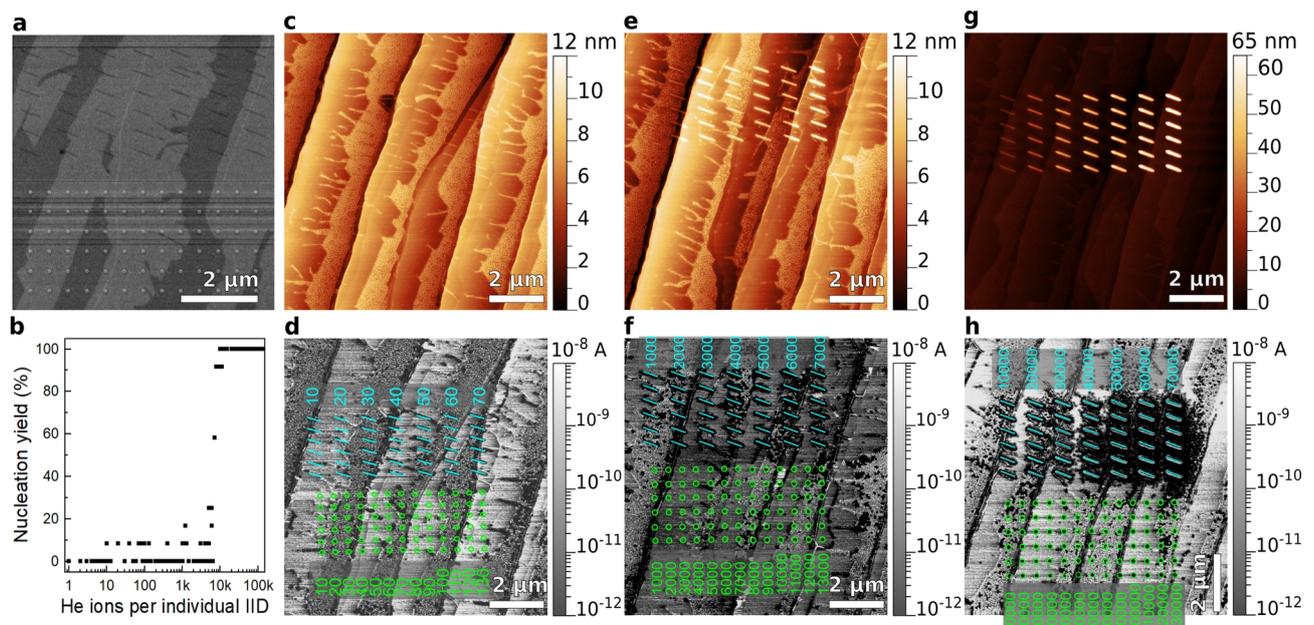

**Fig. S6a**, SEM image of a typical test pattern where in each of the 13 rows, 6 individual IIDs were written 500 nm below each other using between 10000 and 130000 ions per IID. The 7 rows with overlapping IIDs written along lines visible above were written with 10000 to 70000 He ions per IID. **b**, Nucleation yield for h-BN grown for 300 minutes on the individual IIDs. **c-h**, Exemplary test structures written with varying number of ions after the growth of h-BN. **c**, AFM image of test structure written with 10 to 130 ions per individual IID (positions marked by green colored circles) and 10 to 70 ions per overlapping IID (position marked by cyan colored lines). The locations of the IIDs and the number of He ions used per IID in the different rows are noted in the corresponding cAFM image in (**d**). **e**, AFM image and (**f**) corresponding cAFM image of test structure where between 1000 and 13000 He ions were used per individual IID and between 1000 and 7000 He ions per overlapping IID. **g**, AFM image and (**h**) corresponding cAFM image of test structure where between 10000 and 130000 He ions were used per individual IID and between 10000 and 70000 He ions per overlapping IID.

**Supplementary References**